\documentclass[12pt,a4paper,twoside]{article}

\usepackage{array,float,graphicx,amssymb,fancybox}
\usepackage{amsmath,bm,color}
\usepackage{fancyhdr}
\usepackage{lastpage}

\def\T{{\rm T}}

\newcommand{\be}{\begin{equation}}
\newcommand{\ee}{\end{equation}}

\begin{document}

\title{Thermoelasticity and Generalized Thermoelasticity Viewed as Wave Hierarchies}
\author{N. H. Scott\thanks{Email: n.scott@uea.ac.uk}\\
School of Mathematics, University of East Anglia,\\
Norwich Research Park,
 Norwich NR4 7TJ, UK.}
\date{}
\maketitle

\thispagestyle{fancy} 
\lhead{
\emph{IMA Journal of Applied Mathematics} (2008)  {\bf 73}, 123--136.\\ 
doi: 10.1093/imamat/hxm010\\
Published 18 October 2007
}
\rhead{Page\ \thepage\ of\ \pageref{LastPage}}
\cfoot{}



 \begin{abstract}
It is seen how to write the standardÊ form of the four partial differential equations in four unknowns of anisotropic thermoelasticity as a single equation in one variable, in terms of isothermal and isentropic wave operators.  This equation, of diffusive type, is of the eighth order in the space derivatives and seventh order in the time derivatives and so is parabolic in character.  After having seen how to cast the 1D diffusion equation into Whitham's wave hierarchy form it is seen how to recast the full equation, for uni-directional motion, in wave hierarchy form.  The higher order waves are isothermal and the lower order waves are isentropic or purely diffusive.  The wave hierarchy form is then used to derive the main features of the solution of the initial value problem, thereby bypassing the need for an asymptotic analysis of the integral form of the exact solution.   The results are specialized to the isotropic case.  The theory of generalized thermoelasticity associates a relaxation time with the heat flux vector and the resulting system of equations is hyperbolic in character.  It is seen also how to write this system in wave hierarchy form which is again used to derive  the main features of the solution of the initial value problem.  Simpler results are obtained in the isotropic case.

\vspace{1mm}\noindent
\end{abstract}

\section{Introduction and basic equations} 
\setcounter{equation}{0}

King {\em et al}. (1998)  studied the effects of weak hyperbolicity on the diffusion of heat modelled by the equation of telegraphy
\be\label{1.1}\varepsilon\theta_{tt}+\theta_t=\theta_{xx}\ee
in which the temperature increment $\theta(x, t)$ is a function of position $x$ and time $t$.  With all quantities suitably non-dimensionalized, $\varepsilon$ is a dimensionless parameter measuring the ratio of relaxation time scale to diffusion time scale.  For $\varepsilon=0$, equation (\ref{1.1}) reduces to the usual parabolic diffusion equation for heat and it is well known that part of an initially localized disturbance reaches infinity instantaneously.
For $0<\varepsilon \ll 1$, however, equation (\ref{1.1}) is hyperbolic in nature with a small part of the initially localized disturbance propagating out into the undisturbed region by means of waves with large speeds $\pm\sqrt{1/\varepsilon}$ which have amplitudes heavily damped, by the exponential factor ${\rm e}^{-t/2\varepsilon}$.
King {\em et al}. (1998) were able to confirm these results using Whitham's (1974) wave hierarchy approach  
and went on to give a full discussion of the asymptotics of an exact solution of an initial value problem for equation (\ref{1.1}) expressed in integral form.

\pagestyle{fancy}
\fancyhead{}
\fancyhead[RO,LE]{Page\ \thepage\ of\ \pageref{LastPage}}
\fancyhead[LO]{Thermoelasticity  as wave heirarchies}
\fancyhead[RE]{N. H. Scott}

In the present paper we extend the wave hierarchy approach of King {\em et al}. (1998) to classical anisotropic thermoelasticity and also to generalized anisotropic thermoelasticity in which there is a relaxation time as with (\ref{1.1}).  We then use the wave hierarchy form  to derive the main features of the solution of the initial value problem. 
The main advantage of the wave hierarchy approach is that it allows the main features of the solution to be deduced without recourse to a complicated asymptotic analysis of the exact solution obtainable by integral transform methods.

The standard form of the equations of anisotropic thermoelasticity consists of four partial differential equations in 
 four unknowns, the three components of displacement together with temperature, each being a function of space and time. Three of these equations are second order in both space and time whilst 
the fourth is second order in space but  only first order in time. This leads to the equations of 
thermoelasticity having an overall diffusive nature. In this paper it is seen how to write these equations 
in an operator form from which it  is deduced that temperature and each component of displacement 
each satisfy  the same partial differential equation which is of eighth order in space and seventh in time.   
For isotropic thermoelasticity this equation reduces to a well known equation which is fourth order in 
space and third order in time.     Having reduced the four equations of thermoelasticity to one equation in a single  unknown it is 
possible to interpret this equation in terms of Whitham's  (1974) wave hierarchy approach. However, this 
approach needs to be modified because of the presence of two space derivatives not associated with 
two corresponding time derivatives and one time derivative not associated with a corresponding space
 derivative, indicative of the underlying diffusive nature of the equation. The higher order wave 
operator is found to be of eighth order and the lower wave operator of seventh order.  Isothermal 
waves are associated with the higher order operator and isentropic waves with the lower order 
operator.  It has been shown elsewhere that the isothermal and isentropic wave speeds interlace, see Scott (1989a), so  
that Whitham's (1974) stability criterion is satisfied. In the (usual) case of small coupling between the elastic and 
thermal effects it is found that the bulk of the disturbance either travels with the lower order (isentropic) wave speeds or diffuses.

We take the equations of thermoelasticity in the standard form of Chadwick (1979):
\be\label{1}
\begin{array}{c}
\tilde{c}^{\phantom{2}}_{ijkl} \partial^{\phantom{2}}_j\partial^{\phantom{2}}_l  u^{\phantom{2}}_{k} - \beta^{\phantom{2}}_{ij} \partial^{\phantom{2}}_j\theta
=  \rho\partial^2_t  u^{\phantom{2}}_i,\\[2mm]
k^{\phantom{2}}_{ij} \partial^{\phantom{2}}_i\partial^{\phantom{2}}_j\theta - T\beta^{\phantom{2}}_{ij} \partial^{\phantom{2}}_j\partial^{\phantom{2}}_t u^{\phantom{2}}_i  =  \rho c \partial^{\phantom{2}}_t\theta,
\end{array}
\ee
in which ${\bf u(x}, t)$ is the particle displacement vector and $\theta({\bf x}, t)$ the temperature increment, both being functions of position $\bf x$ and time $t$.  The time derivative $\partial/\partial t$ is denoted by $\partial_t$ and the space derivative $\partial/\partial x_j$  is denoted by $\partial_j$.  Repeated suffixes are summed over.  All other quantities occurring in (\ref{1}) are constants evaluated in the reference configuration: $T$ is the ambient absolute temperature, $\rho$ the mass density, $c$ the specific heat and $\beta_{ij}, \;k_{ij},\;\tilde{c}_{ijkl}$ are the components of the temperature coefficient of stress, conductivity and isothermal elasticity tensors, respectively.  The last four quantities are defined by Chadwick (1979).  The specific heat is positive and these tensors are positive definite.

With $\partial_{\bf x}$ denoting $\partial/\partial_{\bf x}$ we define the spatial differential operators
\[\tilde{Q}_{ik}\left(\partial_{\bf x}\right) := \tilde{c}_{ijkl} \partial_j\partial_l,\;\;\;
 \beta_i \left(\partial_{\bf x}\right) := \beta_{ij}\partial_j,\;\;\;
k\left(\partial_{\bf x}\right) :=k_{ij} \partial_i\partial_j,
\]
in terms of which (\ref{1}) may be rewritten as
\[
\left\{\tilde{\bf Q}\left(\partial_{\bf x}\right)  - \rho\partial^2_t{\bf I}
\right\}{\bf u} - \mbox{\boldmath{$\beta$}} \left(\partial_{\bf x}\right)\theta 
= {\bf 0},
\]
\[(\rho c)^{-1}T \mbox{\boldmath{$\beta$}} \left(\partial_{\bf x}\right)\partial_t {\bf u} + 
\left\{\partial_t  - (\rho c)^{-1}k\left(\partial_{\bf x}\right)\right\}\theta = 0.
\]
These equations may be written in $4\times4$ matrix differential operator form as
\begin{equation}
\left( \begin{array}{ccc}
 \tilde{\bf Q}(\partial_{\bf x})  - \rho\partial^{2}_{t}{\bf I}
  &\vline&   - \mbox{\boldmath{$\beta$}}(\partial_{\bf x})\\
[1mm]\hline\rule{0mm}{5mm} 
(\rho c)^{-1}T \mbox{\boldmath{$\beta$}}^\T
(\partial_{\bf x})\,\partial_t  &\vline&  \partial_t - (\rho c)^{-1}k(\partial_{\bf x})\end{array} \right)
\left(\begin{array}{c}{\bf u}\\
[1mm]\hline\rule{0mm}{5mm}
\theta
\end{array} \right)
 = 
\left(\begin{array}{c}{\bf 0}\\
[1mm]\hline\rule{0mm}{5mm}
0
\end{array} \right),
\label{2}
\end{equation}
with $^\T$ denoting the transpose. 
We denote the $4\times4$ matrix appearing in (\ref{2}) by $M$, so that
\begin{equation}
M = \left( \begin{array}{ccc}
 \tilde{\bf Q}(\partial_{\bf x})  - \rho\partial^{2}_{t}{\bf I}
  &\vline&   - \mbox{\boldmath{$\beta$}}(\partial_{\bf x})\\
[1mm]\hline\rule{0mm}{5mm}
 (\rho c)^{-1}T \mbox{\boldmath{$\beta$}}^\T
(\partial_{\bf x})\,\partial_t  &\vline&  \partial_t - (\rho c)^{-1}k(\partial_{\bf x})\end{array} \right).
\label{2b}
\end{equation}

Equation (\ref{2}) is preserved upon multiplication of any row by a differential operator and upon adding to any row a differential operator multiple of any other row.  Clearly, we may not divide by a differential operator but, without doing this, we shall see that we may put the system of equations (\ref{2}) into echelon form, or even diagonal form, by means of these permissible row operations alone.  This gives a systematic way of eliminating some of the variables and reducing the number of equations accordingly.

\section{Preliminary results} 
\setcounter{equation}{0}

\subsection{Elimination of $\theta$ in favour of 
$\bf u$}

We now use row 4 of (\ref{2}) in order to eliminate the $ -  \mbox{\boldmath{$\beta$}}$ entries in column 4 of the matrix operator.  First, multiply rows 1, 2, 3  of (\ref{2}) by the {\em diffusion operator}
\begin{equation}
\partial_t - (\rho c)^{-1}k\left(\partial_{\bf x}\right)
\label{2c}
\end{equation}
and then add $\beta_i\times$ row 4 to each row $i$, $i=1, 2, 3$, of (\ref{2})  to give a set of equations of the form (\ref{2}) except that the matrix $M$ occurring is replaced by
\[
\left( \begin{array}{ccc}
\left\{ \partial_t - (\rho c)^{-1}k(\partial_{\bf x})\right\} \left(\tilde{\bf Q}(\partial_{\bf x})  - \rho\partial^2_t{\bf I} \right)  
 + (\rho c)^{-1} \mbox{\boldmath{$\beta$}}(\partial_{\bf x})\otimes \mbox{\boldmath{$\beta$}}(\partial_{\bf x})\,\partial_t
&\vline& {\bf 0}\\
[1mm]\hline\rule{0mm}{5mm}
 (\rho c)^{-1}T \mbox{\boldmath{$\beta$}}^\T(\partial_{\bf x})\, \partial_t  &\vline&  \partial_t - (\rho c)^{-1}k(\partial_{\bf x})
\end{array} \right)
\]
in which $\otimes$ denotes the dyadic product.  The first three equations no longer contain $\theta$ and may be rewritten in terms of the operator
\begin{equation}
 \hat{\bf Q}(\partial_{\bf x}) := \tilde{\bf Q}(\partial_{\bf x}) 
+  (\rho c)^{-1}T \mbox{\boldmath{$\beta$}}(\partial_{\bf x})\otimes \mbox{\boldmath{$\beta$}}(\partial_{\bf x})
\label{2a}
\end{equation}
as
\begin{equation}
\partial_t\left\{\hat{\bf Q}(\partial_{\bf x})  - \rho\partial^2_t{\bf I} \right\}
{\bf u}
 - (\rho c)^{-1}k(\partial_{\bf x})
\left\{\tilde{\bf Q}(\partial_{\bf x})  - \rho\partial^2_t{\bf I} \right\}{\bf u}
= 0
\label{3}
\end{equation}
in which the operators
\[
\tilde{\bf Q}(\partial_{\bf x})  - \rho\partial^2_t{\bf I}\;\; \mbox{  and  }\;\;
\hat{\bf Q}(\partial_{\bf x})  - \rho\partial^2_t{\bf I}
\]
are termed the {\em isothermal} and {\em isentropic wave operators}, respectively.  The operators $\partial_t$ and $ (\rho c)^{-1}k(\partial_{\bf x})$  in (\ref{3}) are components of the diffusion operator (\ref{2c}).

\subsection{Reduction to a single equation in one variable}

If we multiply (\ref{2}) on the left by ${\rm adj\,}M$, the transposed matrix of cofactors of $M$, and use the property
\[({\rm adj\,}M) M = (\det M) I_4,\]
with $I_4$ denoting the $4\times4$ identity matrix, then we see that equations (\ref{2}) are reduced to a diagonal form which shows that they  may be replaced by
\begin{equation}
(\det M) u_i = 0,\;\;i=1, 2, 3, \;\;(\det M) \theta = 0,
\label{4}
\end{equation}
so that $\theta$ and each component of $\bf u$ satisfies the same partial differential equation.  The matrix differential operator ${\rm adj\,}M$ and the scalar differential operator $\det M$ are both well defined as neither involves division by a differential operator.

We need to make (\ref{4}) more convenient by determining an explicit form for the differential operator $\det M$.  The method follows that of Chadwick (1979) in the harmonic plane wave case.  By applying elementary methods to row 4 of $M$ we find that
\[
\det M = 
\left| \begin{array}{ccc}
 \tilde{\bf Q}(\partial_{\bf x})  - \rho\partial^{2}_{t}{\bf I}
  &\vline&   - \mbox{\boldmath{$\beta$}}(\partial_{\bf x})\\
[1mm]\hline\rule{0mm}{5mm}
 (\rho c)^{-1}T \mbox{\boldmath{$\beta$}}^\T
(\partial_{\bf x})\,\partial_t  &\vline&  \partial_t  \end{array}\right|
 + 
\left| \begin{array}{ccc}
 \tilde{\bf Q}(\partial_{\bf x})  - \rho\partial^{2}_{t}{\bf I}
  &\vline&   - \mbox{\boldmath{$\beta$}}(\partial_{\bf x})\\
[1mm]\hline\rule{0mm}{5mm}
{\bf 0} &\vline&   - (\rho c)^{-1}k
(\partial_{\bf x})\end{array}\right|.
\]
In the first determinant we may take out the common factor $\partial_t$ from row 4 and in what remains add $\beta_i\times$ row 4 to row $i$, $i=1, 2, 3,$ to obtain
\[
\det M = 
\partial_t \left| \begin{array}{ccc}
 \hat{\bf Q}(\partial_{\bf x})  - \rho\partial^{2}_{t}{\bf I}
  &\vline&  {\bf 0}\\
[1mm]\hline\rule{0mm}{5mm}
 (\rho c)^{-1}T \mbox{\boldmath{$\beta$}}^\T(\partial_{\bf x})  &\vline&  1 
\end{array}\right|
 + 
\left| \begin{array}{ccc}
 \tilde{\bf Q}(\partial_{\bf x})  - \rho\partial^{2}_{t}{\bf I}
  &\vline&   - \mbox{\boldmath{$\beta$}}(\partial_{\bf x})\\
[1mm]\hline\rule{0mm}{5mm}
{\bf 0} &\vline&   - (\rho c)^{-1}k(\partial_{\bf x})\end{array}\right|,
\]
where (\ref{2a}) has been used.  Thus we see that
\[
\det M = 
\partial_t\det\left( \hat{\bf Q}(\partial_{\bf x})  - \rho\partial^{2}_{t}{\bf I}\right)
 - (\rho c)^{-1}k(\partial_{\bf x})
\det\left( \tilde{\bf Q}(\partial_{\bf x})  - \rho\partial^{2}_{t}{\bf I}\right)
\]
so that equation (\ref{4}) for $\theta$ is
\begin{equation}
(\rho c)^{-1}k(\partial_{\bf x})\det\left\{ \tilde{\bf Q}(\partial_{\bf x})  - \rho\partial^{2}_{t}{\bf I}\right\}\theta
 - \partial_t\det\left\{ \hat{\bf Q}(\partial_{\bf x})  - \rho\partial^{2}_{t}{\bf I}\right\}\theta
  = 0.
\label{5}
\end{equation}
Each component of $\bf u$ also satisfies an equation of the form (\ref{5}), with $\theta$ replaced by $u_i$.  The first  operator in (\ref{5}) is of eighth order in the space derivatives and sixth order in the time derivative and the second is sixth order in the space derivatives and seventh in the time derivative, so that, overall, (\ref{5}) is eighth order in the space derivatives and seventh order in the time derivative.

It is possible to write (\ref{5}) in terms of one determinant only but at the expense of introducing two further copies of the diffusion operator (\ref{2c}).  Using definition (\ref{2a}) and the identity
\begin{equation}
\det({\bf A} + \alpha{\bf a\otimes a}) = \det{\bf A} + \alpha{\bf a}\cdot({\rm    adj}\, {\bf A)a}
\label{7}
\end{equation}
(valid for all scalars $\alpha$, vectors $\bf a$ and tensors
 $\bf A$)
in the second operator of (\ref{5}) we see that (\ref{5}) can be rewritten as
\begin{equation}\label{2.7}
\left\{\left(\partial_t-(\rho c)^{-1}k\right)
\det(\tilde{\bf Q}-\rho\partial^2_t{\bf I}) 
+ (\rho c)^{-1}T\partial_t\,  \mbox{\boldmath{$\beta$}}
\cdot {\rm adj\,}(\tilde{\bf Q} -\rho\partial^2_t{\bf I})\mbox{\boldmath{$\beta$}}
\right\}\theta = 0,
\end{equation}
suppressing temporarily the dependence on $\partial_{\bf x}$. 
On multiplying by $(\partial_t-(\rho c)^{-1}k)^2$ and using (\ref{7}) and (\ref{2a}) again we are eventually able to cast (\ref{5}) into the form
\begin{equation}
\det\left[(\rho c)^{-1}k(\partial_{\bf x})\left\{ \tilde{\bf Q}(\partial_{\bf x})  - \rho\partial^{2}_{t}{\bf I}\right\}
 - \partial_t\left\{ \hat{\bf Q}(\partial_{\bf x})  - \rho\partial^{2}_{t}{\bf I}\right\}\right]\theta
  = 0
\label{8}
\end{equation}
involving a single determinant.  However, (\ref{8}) is of higher order than (\ref{5}), being of twelfth order in the space derivatives and ninth order in the time derivative.

\subsection{Diagonalization in terms of one space variable}

Although the wave operators $\tilde{\bf Q}(\partial_{\bf x})$  and $\hat{\bf Q}(\partial_{\bf x})$ are real and symmetric it is not usually possible to diagonalize them as the rotations required would involve division by differential operators.  However, in the case of spatial dependence upon one coordinate only we now see that such diagonalization is indeed possible.  Let the space dependence be through $x={\bf n\cdot x}$ only, where $\bf n$ is a given real unit vector denoting the direction of wave propagation.  Then, with $\partial^2_x$ denoting $\partial^2/\partial x^2$, 
\[
\tilde{\bf Q}(\partial_{\bf x}) = \tilde{\bf Q}({\bf n})\partial^2_x,\qquad
\hat{\bf Q}(\partial_{\bf x}) = \hat{\bf Q}({\bf n})\partial^2_x,\qquad
k(\partial_{\bf x}) = k({\bf n})\partial^2_x,
\]
where the {\em isothermal} and {\em isentropic acoustic tensors } are defined in components by
\[\tilde{Q}_{ik}({\bf n}) = \tilde{c}_{ijkl}n_jn_l,\qquad
\hat{ Q}_{ik}({\bf n}) = \tilde{ Q}_{ik}({\bf n})
 + (\rho c)^{-1}T\,\beta_{ij}n_j\,\beta_{kl}n_l,
\]
respectively, and $k({\bf n})=k_{ij}n_in_j$ is the thermal conductivity in the direction $\bf n$.  Then (\ref{5}) may be rewritten as
\begin{equation}
(\rho c)^{-1}k({\bf n})\partial^2_x\det\left\{{\bf I}\partial^2_t - \rho^{-1} \tilde{\bf Q}({\bf n})\partial^2_x\right\}\theta
 - \partial_t\det\left\{{\bf I}\partial^2_t - \rho^{-1} \hat{\bf Q}({\bf n})\partial^2_x\right\}\theta = 0.
\label{9}
\end{equation}

The eigenvalues of the real, symmetric matrices $\rho^{-1}\tilde{\bf Q}({\bf n})$ and $\rho^{-1}\hat{\bf Q}({\bf n})$ are denoted by $\tilde{c}^2_i$ and $\hat{c}^2_i$, $i=1, 2, 3, $ ordered according to
\begin{equation}
\tilde{c}^2_1<\tilde{c}^2_2<\tilde{c}^2_3,\qquad
\hat{c}^2_1<\hat{c}^2_2<
\hat{c}^2_3.
\label{10}
\end{equation}
These are the isothermal and isentropic squared wave speeds, respectively, guaranteed positive if the strong ellipticity of $\tilde{\bf c}$ holds.  They depend upon $\bf n$ and are assumed to be distinct.  Each of $\tilde{\bf Q}({\bf n})$ and $\hat{\bf Q}({\bf n})$ may be diagonalized and the determinants in (\ref{9}) evaluated to give
\begin{equation}
(\rho c)^{-1}k\,\partial^2_x\prod_{i=1}^3\left(\partial^2_t - \tilde{c}^2_i\partial^2_x\right)\theta
 - \partial_t\prod_{i=1}^3\left(\partial^2_t - \hat{c}^2_i\partial^2_x\right)\theta = 0,
\label{11}
\end{equation}
where $k$ is written for $k({\bf n})$. 
Of course, $\theta$ in (\ref{11}) may be replaced by any component of $\bf u$.  This equation is in Whitham's wave hierarchy form, see Whitham (1974), except for the separated appearance of the components $\partial_t$ and $ (\rho c)^{-1}k({\bf n})\partial^2_x$ of the diffusion operator (\ref{2c}).  
The first term of (\ref{11}) represents the higher order waves and the second the lower order waves in Whitham's wave hierarchy. 

\subsection{The diffusion equation as a wave hierarchy}

In the special case where the temperature coefficient of stress vanishes, i.e. $\beta_{ij}=0$, the equations (\ref{1}) of thermoelasticity decouple into three purely elastic wave equations for the displacement components and a separate diffusion equation
for the temperature:
\begin{equation}
(\rho c)^{-1}k\,\partial^2_x\theta - \partial_t\theta = 0,
\label{12a}
\end{equation}
in which the space variable is $x={\bf n\cdot x}$, as before.     This diffusion equation may be put 
into wave hierarchy form by observing that it may be obtained in the limit $\epsilon\rightarrow 0,\;\;\delta\rightarrow 0$ from
\begin{equation}
\epsilon^2\left(\partial^2_t-\frac{k}{\rho c\epsilon^2}\partial^2_x\right)\theta 
 + \left(\partial_t+\delta\partial_x\right)\theta = 0.
\label{12}
\end{equation}
There are two higher order waves with speeds $\pm \sqrt{k/\rho c\epsilon^2}$ and one lower order wave with speed $-\delta$, which, for $\epsilon$ and $\delta$ small enough, satisfy Whitham's (1974) stability criterion
\[- \sqrt{k/\rho c\epsilon^2} < -\delta < \sqrt{k/\rho c\epsilon^2}.\]
In the limit $\epsilon\rightarrow 0,\;\;\delta\rightarrow 0$ of the diffusion equation the higher order wave speeds become infinite whilst the lower order wave speed vanishes.

\vspace{1mm}\noindent{\em Higher order waves.}\hspace{2mm}To follow a wave moving to the right with speed $\sqrt{k/\rho c\epsilon^2}$ we approximate $\partial_t\approx - \sqrt{k/\rho c\epsilon^2}\,\partial_x$ everywhere in (\ref{12}) except in the wave operator $\partial_t +  \sqrt{k/\rho c\epsilon^2}\,\partial_x$ itself to obtain
\[
\epsilon^2\left(-2\sqrt{k/\rho c}\right)\left(\partial_t + \sqrt{k/\rho c\epsilon^2} \,\partial_x\right)\partial_x\theta + \left(-\sqrt{k/\rho c} + \epsilon\delta\right) \partial_x\theta=0.
\]
We let $\delta\rightarrow 0$ and ignore the overall factor $\partial_x$ as it corresponds to the remnants of other waves to obtain
\[
\left(\partial_t + \sqrt{k/\rho c\epsilon^2}\partial_x\right)\theta 
 + \frac{1}{2\epsilon^2}\,\theta = 0,
\]
which has general solution
\[\theta(x, t) = f(x -  \sqrt{k/\rho c\epsilon^2}\,t){\rm e}^{-t/2\epsilon^2},
\]
where $\theta(x, 0) = f(x)$ is the arbitrary initial profile of the disturbance. 
As $\epsilon\rightarrow 0$ we are left with $\theta\rightarrow 0$ for all $t>0$.  Thus no disturbance at all travels with the higher order waves in the limit $\epsilon\rightarrow 0$ required by the diffusion equation.

\vspace{1mm}\noindent{\em Lower order wave.}\hspace{2mm}To follow the wave moving with speed $-\delta$ we approximate $\partial_t\approx-\delta\partial_x$ everywhere in (\ref{12}) except in the lower order wave operator $\partial_t + \delta\partial_x$ to obtain
\[
\left(\epsilon\delta+\sqrt{k/\rho c}\right)\left(\epsilon\delta-\sqrt{k/\rho c}\right) \partial^2_x\theta + \left(\partial_t+\delta\partial_x\right)\theta = 0.
\]
In the limit $\epsilon\rightarrow 0,\;\;\delta\rightarrow 0$ this becomes the original diffusion equation (\ref{12a}).  Thus, not surprisingly, all the disturbance goes with the lower order wave which, since the speed is zero, has here degenerated into a diffusion operator.

\section{Thermoelasticity as a wave hierarchy}
\setcounter{equation}{0}

Guided by the results for the diffusion equation (\ref{12a}) we may put the equation of thermoelasticity (\ref{11}) into wave hierarchy form by observing that it may be obtained in the limit $\epsilon\rightarrow 0,\;\;\delta\rightarrow 0$ from
\begin{equation}
\epsilon^2\left(\partial^2_t-\frac{k}{\rho c\epsilon^2} \partial^2_x\right)\prod_{i=1}^3\left(\partial^2_t - \tilde{c}^2_i\partial^2_x\right)\theta
  +  \left(\partial_t+\delta\partial_x\right) \prod_{i=1}^3\left(\partial^2_t - \hat{c}^2_i\partial^2_x\right)\theta = 0.
\label{3.1}
\end{equation}
There are eight higher order waves with speeds $\pm \sqrt{k/\rho c\epsilon^2}$, $\pm \tilde{c}_i$, $i=1, 2, 3$, and seven lower order waves with speeds $-\delta$, $\pm \hat{c}_i$, $i=1, 2, 3$, for which Whitham's stability criterion is
\begin{eqnarray}
  -  \sqrt{k/\rho c\epsilon^2}  <-\hat{c}_3<-\tilde{c}_3<-\hat{c}_2<-\tilde{c}_2<-\hat{c}_1 <-\tilde{c}_1 \nonumber\\ 
  < -\delta<\tilde{c}_1<\hat{c}_1<\tilde{c}_2<\hat{c}_2<\tilde{c}_3 <\hat{c}_3<\sqrt{k/\rho c\epsilon^2}.
\label{3.2}
\end{eqnarray}
It has already been shown, see Scott (1989a), that the interlacing of the isothermal and isentropic wave speeds, as in (\ref{3.2}), is a necessary and sufficient condition for the  stability of thermoelastic waves and so the wave hierarchy approach confirms this result.   As $\epsilon\rightarrow 0,\;\;\delta\rightarrow 0$  the outer and middle inequalities of (\ref{3.2}) clearly hold as in the case of the diffusion equation (\ref{12}).  Also, as in that case we find that the disturbance travelling with the higher order speeds $\pm  \sqrt{k/\rho c\epsilon^2} $ vanishes as $\epsilon\rightarrow 0$ for all $t>0$ and that the lower order wave with speed $-\delta$ degenerates to a diffusion operator as $\delta\rightarrow 0$.  Therefore we may take the limits $\epsilon\rightarrow 0,\;\;\delta\rightarrow 0$  in (\ref{3.1}) and revert, from now on, to the original equation (\ref{11}) of thermoelasticity.

\vspace{1mm}\noindent{\em Higher order waves.}\hspace{2mm}Let us follow the higher order wave moving with speed $\tilde{c}_i$, so that we may approximate $\partial_t\approx-\tilde{c}_i\partial_x$ except in the operator $\partial_t + \tilde{c}_i\partial_x$.  Ignoring the residual wave operator $\partial^7_x$  we find that (\ref{11}), applied now to a displacement component $u_1$, reduces to
\be\label{3.3}
\left(\partial_t + \tilde{c}_i\partial_x\right)u_1 + \eta_i\, u_1=0, \quad i=1, 2, 3,
\ee
where $\eta_i$ is given by
\be\label{3.4}
\eta_i= -\, \frac{\rho c}{2k}\cdot\frac{d(\tilde{c}^2_i)}{e^\prime(\tilde{c}^2_i)}
\ee
in which the cubic polynomials $d$ and $e$ are defined by
\[d(v^2):=\prod_{i=1}^{i=3}(v^2-\hat{c}^2_i),\qquad e(v^2):=\prod_{i=1}^{i=3}(v^2-\tilde{c}^2_i),
\]
and prime denotes differentiation with respect to argument.  The constants $\eta_i$ are positive because of (\ref{3.2}).
The general solution of (\ref{3.3}) is
\be\label{3.5}
u_1=f(x-\tilde{c}_it){\rm e}^{-\eta_i t}
\ee
with $f(x)$ the arbitrary initial profile.  This is a permanent-form travelling wave with exponential damping.  If there is only weak coupling between mechanical and thermal effects then $\hat{c}^2_i/\tilde{c}^2_i - 1$ is small and so $\eta_i$ is small leading to light damping.  If, on the other hand, the material is virtually a non-conductor then $k\rightarrow 0$ so that $\eta_i\rightarrow\infty$ and the higher order waves are heavily damped.

\vspace{1mm}\noindent{\em Lower order waves.}\hspace{2mm}We now follow a lower order wave moving with speed $\hat{c}_i$, so that we may approximate $\partial_t\approx-\hat{c}_i\partial_x$ except in the operator $\partial_t + \hat{c}_i\partial_x$ itself.  Ignoring the residual wave operator $\partial^6_x$  we find that (\ref{11}) reduces to
\be\label{3.6}
\left(\partial_t + \hat{c}_i\partial_x\right)u_1 = D_i\partial^2_x u_1,\quad i=1, 2, 3,
\ee
a convected diffusion equation with diffusivity
\be\label{3.7}
D_i = \frac{k}{2\rho c}\cdot \frac{e(\hat{c}^2_i)}{\hat{c}^2_i d^\prime(\hat{c}^2_i)},
\ee
positive on account of (\ref{3.2}).  For each $i=1, 2, 3$, the solution of (\ref{3.6}) may be written as the convolution
\be\label{3.8}
u_1(x, t)=\int_{-\infty}^\infty f(\xi)
\delta_\varepsilon(x-\hat{c}_it-\xi)\,d\xi
\ee
of the initial profile $u_1(x, 0)=f(x)$ with the function
\[\delta_\varepsilon(X):=\frac{1}{\sqrt{\pi\varepsilon}}
\exp\left(-\frac{X^2}{\varepsilon}\right),\qquad \varepsilon:=4D_i t,
\]
which furnishes a $\delta$-function sequence in the sense that
\[\lim_{\varepsilon\to 0}\delta_\varepsilon(X) = \delta(X),
\]
the Dirac $\delta$-function.  As $t\to 0$ we see from the properties of the $\delta$-function that the initial profile is recovered from (\ref{3.8}).  If the initial disturbance is a spike $f(x)=\delta(x)$ we see from  (\ref{3.8}) that the subsequent disturbance is a spike moving with speed $\hat{c}_i$ but broadening by diffusion over a length scale $(D_it)^{1/2}$.
In the case of either small thermoelastic coupling or small conductivity we may regard $D_i$ as small so that, by similar reasoning, the lower order wave is a travelling wave of permanent form $f(x-\hat{c}_it)$ modified by diffusion over the same length scale $(D_it)^{1/2}$.

To study the degenerate lower order wave operator $\partial_t$ in (\ref{11}) we replace it by zero everywhere else to obtain the diffusion equation 
\be\label{3.9}
\frac{k}{\rho c}\cdot\frac{\tilde{c}^2_1\tilde{c}^2_2\tilde{c}^2_3}
{\hat{c}^2_1\hat{c}^2_2\hat{c}^2_3}\,\partial^2_x\theta
 - \partial_t\theta = 0,
\ee
where an operator $\partial^6_x$ has been ignored.  Thus the temperature in coupled thermoelasticity diffuses with a diffusivity less than that of (\ref{12a}) on account of (\ref{3.2}).

\subsection{The isotropic case}
For an isotropic material we have  
$\beta_{ij}=\beta\delta_{ij}, \;k_{ij}=k\delta_{ij},\;\tilde{c}_{ijkl} = 
\lambda\delta_{ij}\delta_{kl} + \mu
(\delta_{ik}\delta_{jl} + \delta_{il}\delta_{jk})$,
in terms of the components of the unit tensor $\delta_{ij}$ and the Lam\'{e} moduli $\lambda$ and $\mu$.
Taking the $x_3$-direction parallel to the direction $\bf n$ of wave propagation, we 
find that the wave speeds are given by
\be\label{3.10}
\tilde{c}^2_1=\hat{c}^2_1=\tilde{c}^2_2=\hat{c}^2_2=
\mu/\rho,\quad \tilde{c}^2_3 = (\lambda+2\mu)/\rho,\quad
\hat{c}^2_3 = \tilde{c}^2_3 + T\beta^2/\rho^2c.
\ee
As is well known, the transverse displacements $u_1$ and $u_2$  are  purely elastic in character and independent of temperature effects.  In fact, they satisfy the isothermal wave equation
$(\partial_t^2 - \tilde{c}^2_1\partial_x^2)u_{1,2}=0$.
This wave operator occurs twice in (\ref{11}) when specialized to the isotropic case and on removing these repeated factors (\ref{11}) reduces to the following equation for $\theta$:
\begin{equation}
(\rho c)^{-1}k\,\partial^2_x\left(\partial^2_t - \tilde{c}^2_3\partial^2_x\right)\theta
 - \partial^{\phantom{2}}_t\left(\partial^2_t - \hat{c}^2_3\partial^2_x\right)\theta = 0.
\label{3.11}
\end{equation}
This equation remains valid if $\theta$ is replaced by the longitudinal displacement $u_3$.
Iannece \& Starita (1988) discuss (\ref{3.11}), with a different scaling, in the context of the thermomechanics of fluids.

By direct calculation, or by specializing the results of the previous subsection, the equations of disturbances propagating with the higher order wave speed $\tilde{c}_3$, with the lower order wave speed $\hat{c}_3$, and as the lower order degenerate diffusion operator are
\be\label{3.12}
\begin{array}{l}
\left(\partial_t + \tilde{c}_3\partial_x\right)u_1 + \displaystyle\frac{\rho c}{2k}(\hat{c}^2_3 - \tilde{c}^2_3)\, u_1=0, \\[3mm]
\left(\partial_t + \hat{c}_3\partial_x\right)u_1 = 
\displaystyle\frac{k}{2\rho c}\cdot\frac{\hat{c}^2_3 - \tilde{c}^2_3}{\hat{c}^2_3}\partial^2_xu_1 ,\\[3mm]
\displaystyle\frac{k}{\rho c}\cdot\frac{\tilde{c}^2_3}
{\hat{c}^2_3}\,\partial^2_x\theta
 - \partial_t\theta = 0,
\end{array}
\ee
corresponding to (\ref{3.3}), (\ref{3.6}) and (\ref{3.9}), respectively.  Leslie \& Scott (1998) investigated by other means the stability of longitudinal waves of sinusoidal form in isotropic thermoelasticity and reached the same conclusions on stability.


\section{Generalized thermoelasticity as a wave hierarchy} 
\setcounter{equation}{0}

\subsection{Preliminary results} 

Generalized thermoelasticity associates a relaxation time $\tau$ with the heat flux vector $\bf q$ so that
\[q_i+\tau \partial_tq_i = -k_{ij}\partial_j\theta,\]
in which $k_{ij}$ are the thermal conductivity components and $\theta$ is temperature, as before.  If $\tau=0$ this equation reverts to Fourier's law of heat conduction and the classical theory of thermoelasticity results as discussed in this paper until now.  With $\tau>0$ it is found that the equations of generalized thermoelasticity are the same as those of classical thermoelasticity (\ref{1}) except that in the last of these equations the operator $\partial_t$ is replaced by $\Delta_t$, defined by
\begin{equation}
\Delta_t := \partial_t + \tau \partial_t^2,
\label{4.0}
\end{equation}
leading to
\be\label{4.1}
\begin{array}{c}
\tilde{c}_{ijkl} \partial_j\partial_l  u_{k} - \beta_{ij} \partial_j\theta
=  \rho\partial^2_t  u_i,\\[2mm]
k_{ij} \partial_i\partial_j\theta - T\beta_{ij} \partial_j\Delta_t u_i  =  \rho c \Delta_t\theta
\end{array}
\end{equation}
in place of (\ref{1}), see Scott (1989b) and Leslie \& Scott (2004).  This has the effect of replacing $\partial_t$ by $\Delta_t$ in the fourth row of the matrix $M$ defined by (\ref{2b}), so that the result of eliminating $\theta$ in favour of $\bf u$ is now 
\begin{equation}
\Delta_t\left\{\hat{\bf Q}(\partial_{\bf x})  - \rho\partial^2_t{\bf I} \right\}
{\bf u}
 - (\rho c)^{-1}k(\partial_{\bf x})
\left\{\tilde{\bf Q}(\partial_{\bf x})  - \rho\partial^2_t{\bf I} \right\}{\bf u}
= 0
\label{4.2}
\end{equation}
in place of (\ref{3}) in the classical case.

In generalized thermoelasticity the reduction of (\ref{4.1}) to a single equation in one variable gives
\begin{equation}
(\rho c)^{-1}k(\partial_{\bf x})\det\left\{ \tilde{\bf Q}(\partial_{\bf x})  - \rho\partial^{2}_{t}{\bf I}\right\}\theta
 - 
\Delta_t\det\left\{ \hat{\bf Q}(\partial_{\bf x})  - \rho\partial^{2}_{t}{\bf I}\right\}\theta
  = 0
\label{4.3}
\end{equation}
in place of  (\ref{5}) in the classical case. 
Each component of $\bf u$ also satisfies an equation of the form (\ref{4.3}), with $\theta$ replaced by $u_i$.  The first  operator in (\ref{4.3}) is of eighth order in the space derivatives and sixth order in the time derivative and the second is sixth order in the space derivatives and eighth in the time derivative, so that, overall, (\ref{4.3}) is eighth order in both the space and time derivatives. This is in contrast with (\ref{5}) which is only seventh order in the time derivative.

Using the methods of Section 2.2 it is possible to rewrite (\ref{4.3}) as
\begin{equation}
\det\left[(\rho c)^{-1}k(\partial_{\bf x})\left\{ \tilde{\bf Q}(\partial_{\bf x})  - \rho\partial^{2}_{t}{\bf I}\right\}
 - 
\Delta_t\left\{ \hat{\bf Q}(\partial_{\bf x})  - \rho\partial^{2}_{t}{\bf I}\right\}\right]\theta
  = 0
\label{4.4}
\end{equation}
involving a single determinant, which is to be compared with (\ref{8}) in the classical case.  However, (\ref{4.4}) is of higher order than (\ref{4.3}), being of twelfth order in both the space and time derivatives.  This contrasts with (\ref{8}) which is of only ninth order in the time derivative.

In terms of the single space variable $x={\bf n}\cdot{\bf x}$, (\ref{4.3}) reduces to
\begin{equation}
(\rho c)^{-1}k\,\partial^2_x\prod_{i=1}^3\left(\partial^2_t - \tilde{c}^2_i\partial^2_x\right)\theta
 - 
( \partial_t + \tau \partial_t^2)\prod_{i=1}^3\left(\partial^2_t - \hat{c}^2_i\partial^2_x\right)\theta = 0,
\label{4.5}
\end{equation}
in which the occurrence of $\Delta_t$ has been made explicit through (\ref{4.0}).  Equation (\ref{4.5}) reduces to 
(\ref{11}) in the classical case upon taking $\tau=0$. 
Of course, $\theta$ in (\ref{4.5}) may be replaced by any component of $\bf u$.  We remember that the thermal conductivity $k$ and the wave speeds $\tilde{c}^2_i$ and $\hat{c}^2_i$ depend on the direction of wave propagation $\bf n$.

\subsection{The wave hierarchy}

The first term of (\ref{4.5}) is of eighth order in $\partial_x$ and sixth in $\partial_t$ and is part of the higher order wave operator.  The new term proportional to $\tau$ is of sixth order in $\partial_x$ and eighth in $\partial_t$ and so must be viewed as part of the higher order wave operator.  The remaining term is the lower order wave operator, of sixth order in $\partial_x$ and seventh in $\partial_t$, and is the same as that in the classical case (\ref{11}).  It is helpful therefore to rewrite (\ref{4.5}) in the form
\begin{equation}\label{4.7}
\tau \partial_t^2\prod_{i=1}^3\left(\partial^2_t - \hat{c}^2_i\partial^2_x\right)\theta
- (\rho c)^{-1}k\,\partial^2_x\prod_{i=1}^3\left(\partial^2_t - \tilde{c}^2_i\partial^2_x\right)\theta
 +
 \partial_t\prod_{i=1}^3\left(\partial^2_t - \hat{c}^2_i\partial^2_x\right)\theta   
= 0,
\end{equation}
the first two terms of which constitute the higher order wave operator and the last term is the lower order wave operator.
Taken together, these first two terms are of eighth order in both 
$\partial_x$ and  $\partial_t$ with only even powers of each occurring.  It is therefore natural to ask if we can combine these two terms into a single higher order wave operator of the form
\[
\tau \partial_t^2\prod_{i=1}^3\left(\partial^2_t - \hat{c}^2_i\partial^2_x\right)
- (\rho c)^{-1}k\,\partial^2_x\prod_{i=1}^3\left(\partial^2_t - \tilde{c}^2_i\partial^2_x\right)
\equiv
\tau \prod_{i=1}^4\left(\partial^2_t - \overline{c}^2_i\partial^2_x\right)
\]
for some real and positive wave speeds $\overline{c}_i$, $i=1, 2, 3, 4$, to be determined.  By allowing both sides of this operator to act on the permanent-form wave $f(x-vt)$ travelling with speed $v$, we see that the squared speeds $\overline{c}^2_i$, $i=1, 2, 3, 4$, are determined as the roots of the following quartic equation in $v^2$:
\be\label{4.8}
h(v^2) :=g(v^2)-(\rho c\tau)^{-1}kj(v^2) = 0,
\ee
where the cubic polynomial $j$ and the quartic polynomial $g$ are defined by
\[j(v^2):= \prod_{i=1}^3\left(v^2 - \tilde{c}^2_i\right)
\quad\mbox{and}\quad
g(v^2):=v^2\prod_{i=1}^3\left(v^2 - \hat{c}^2_i\right).\]
By examining the sign changes of the polynomial $h$ it can be shown that it has four real and positive
zeros, which we may denote by $\overline{c}^2_i$, $i=1, 2, 3, 4$, and that they may be ordered according to
\be\label{4.9}
0<\overline{c}^2_1<\tilde{c}^2_1<\hat{c}^2_1<\overline{c}^2_2<\tilde{c}^2_2<\hat{c}^2_2<\overline{c}^2_3<\tilde{c}^2_3<\hat{c}^2_3<\overline{c}^2_4.
\ee
Then $h$ can be factorized into
\[h(v^2) \equiv \prod_{i=1}^4\left(v^2 - \overline{c}^2_i\right).\]

This real factorization of $h$ implies that the equation (\ref{4.7}) of generalized thermoelasticity can be written in wave-hierarchy form as
\begin{equation}\label{4.10}
\tau  \prod_{i=1}^4\left(\partial^2_t - \overline{c}^2_i\partial^2_x\right)\theta
 +
 \partial_t\prod_{i=1}^3\left(\partial^2_t - \hat{c}^2_i\partial^2_x\right)\theta   
= 0,
\end{equation}
with $\pm\overline{c}_i$, $i=1, 2, 3, 4$,  denoting the higher order wave speeds and $\pm\hat{c}_i$, $i=1, 2, 3$,  the lower order wave speeds.  (Once again, the lower order wave corresponding to the lone operator $\partial_t$ has zero wave speed and 
degenerates into a diffusion process).  Since $\tau>0$ and the interlacing properties (\ref{4.9}) hold, Whitham's (1974) stability criterion holds and the wave-hierarchy equation (\ref{4.10}) of generalized thermoelasticity is seen to be stable.

It is necessary to examine the monotonicity properties of the zeros $\overline{c}^2_i$ of $h$ regarded as functions of $\tau$. 
It is clear from (\ref{4.8}) that as $\tau\to 0$, $\overline{c}^2_i \to \tilde{c}^2_i$, $i=1, 2, 3$, 
and $\overline{c}^2_4 \to \infty$, and that as $\tau\to \infty$, $\overline{c}^2_i \to \hat{c}^2_{i-1}$, 
$i=1, 2, 3, 4$, defining $\hat{c}^2_0\equiv 0$ for convenience.  
The zeros $\overline{c}^2_i$ of $h$ may be expanded for small $\tau$ as
\be\label{4.11}
\begin{array}{rcl}
\overline{c}^2_i &=& \displaystyle\tilde{c}^2_i + \frac{\rho c\tau}{k}\cdot
\frac{g(\tilde{c}^2_i)}{j^\prime(\tilde{c}^2_i)} + O(\tau^2),\;\; i=1, 2, 3,\\[2mm] 
\overline{c}^2_4 &=&\displaystyle \frac{k}{\rho c\tau} +
\hat{c}^2_1 +\hat{c}^2_2+\hat{c}^2_3 - \tilde{c}^2_1 - \tilde{c}^2_2 - \tilde{c}^2_3 +O(\tau),
\end{array}
\ee
where prime denotes differentiation with respect to argument.
From (\ref{3.2}), the quantity $g(\tilde{c}^2_i)/j^\prime(\tilde{c}^2_i)$, $i=1, 2, 3$, is negative
so that, for small $\tau$, each function $\overline{c}^2_i$ is monotonic decreasing in $\tau$.
For large $\tau$ we find that
\[\overline{c}^2_i = \hat{c}^2_{i-1} + \frac{k}{\rho c\tau}\cdot
\frac{j(\hat{c}^2_{i-1})}{g^\prime(\hat{c}^2_{i-1})} + O(\tau^{-2}),\quad i=1, 2, 3, 4,
\]
remembering that $\hat{c}^2_0\equiv 0$.
From (\ref{3.2}), we see that $j(\hat{c}^2_{i-1})/g^\prime(\hat{c}^2_{i-1})$, $i=1, 2, 3, 4$, is positive, so that
for large $\tau$, $\overline{c}^2_i$ is monotonic decreasing in $\tau$.
In fact, we can show further that $\overline{c}^2_i$ is monotonic decreasing on the whole of the $\tau$-range
by arguing as follows.  First note that the only dependence of $h$ upon $\tau$ is that explicit in (\ref{4.8}).
Denote by $h^+$ the quartic polynomial $h$ evaluated at $\tau^+$, selected such that $\tau^+>\tau$, and eliminate $j(v^2)$ to obtain
\[\tau^+h^+=\tau\prod_{i=1}^4\left(v^2 - \overline{c}^2_i\right) + 
(\tau^+-\tau)v^2\prod_{i=1}^3\left(v^2 - \hat{c}^2_i\right).
\]
Denoting the zeros of $h^+$ by $\overline{c}^2_i(\tau^+)$, $i=1, 2, 3, 4$, an examination of the sign changes of the right hand side of this equation shows that
\[\hat{c}^2_{i-1}<\overline{c}^2_i(\tau^+)<\overline{c}^2_i,\quad i=1, 2, 3, 4,
\]
so that the zeros $\overline{c}^2_i$ of $h$ are monotonically decreasing functions of $\tau$.

We now use the results of the previous paragraph to analyse the properties of the higher and lower order waves.

\vspace{1mm}\noindent{\em Higher order waves.}\hspace{2mm}Let us follow the higher order wave moving with speed $\overline{c}_i$, $i=1, 2, 3, 4$, so that we may approximate $\partial_t\approx-\overline{c}_i\partial_x$ except in the operator $\partial_t + \overline{c}_i\partial_x$.  Ignoring the residual wave operator $\partial^7_x$  we find that (\ref{4.10}), applied now to a displacement component $u_1$, reduces to
\be\label{4.12}
\left(\partial_t + \overline{c}_i\partial_x\right)u_1 + \eta_i u_1=0,\quad i=1, 2, 3, 4,
\ee
where the constant $\eta_i$ is given by
\be\label{4.13}
\eta_i= \frac{g(\overline{c}^2_i)}
{2\tau\overline{c}^2_i h^\prime(\overline{c}^2_i)},
\ee
positive because of (\ref{4.9}).
The general solution of (\ref{4.12}) is of the form (\ref{3.5}) with $\tilde{c}_i$ replaced by $\overline{c}_i$ and so is a permanent-form travelling wave with exponential damping.  As $\tau\to 0$ we find from (\ref{4.11}) that 
$\overline{c}^2_i\to\tilde{c}^2_i$, $i=1, 2, 3$, and $\tau\overline{c}^2_4\to k/\rho c$, so that (\ref{4.12}) and (\ref{4.13}) become, respectively, (\ref{3.3}) and (\ref{3.4}), the corresponding equations for $\tau=0$, as we would expect.  The fourth wave, which moves with speed $\overline{c}_4$, behaves differently.  For small $\tau$, equation (\ref{4.12}) in the case $i=4$ reduces approximately to
\be\label{4.14}
\left(\partial_t + \overline{c}_4\partial_x\right)u_1 + \frac{1}{2\tau} u_1=0\quad\mbox{in which}\quad
\overline{c}_4=\sqrt{\left(\frac{k}{\rho c\tau}\right)}
\ee
with general solution $u_1=f(x-\overline{c}_4t){\rm e}^{-t/2\tau}$.  For small $\tau$, this is a wave of permanent form that is fast moving and heavily damped.  Thus three of the waves propagating in the case of small relaxation time $\tau$ are well approximated by their counterparts in the classical theory of thermoelasticity ($\tau=0$), whilst the fourth, with speed $\overline{c}_4$, behaves quite differently in that it moves very rapidly and carries very little of the initial disturbance with it.  This situation is the same as that described by King {\em et al.} (1998) in the context of the equation of telegraphy in which a small relaxation time in the diffusion process induces weak hyperbolicity.  In fact, equation (\ref{4.14}) here is the same as King {\em et al.} (1998, equation (1.15)) with a different scaling.

\vspace{1mm}\noindent{\em Lower order waves.}\hspace{2mm}Following the lower order wave with speed $\hat{c}_i$,  we may approximate $\partial_t\approx-\hat{c}_i\partial_x$ except in the operator $\partial_t + \hat{c}_i\partial_x$.  Ignoring the residual wave operator $\partial^6_x$  we find that (\ref{4.10}) reduces to the convected diffusion equation
\be\label{4.15}
\left(\partial_t + \hat{c}_i\partial_x\right)u_1 = D_i\partial^2_x u_1,\quad i=1, 2, 3,
\ee 
where
\be\label{4.16}
D_i = \frac{- \tau h(\hat{c}^2_i)}
{2g^\prime(\hat{c}^2_i)} 
\ee
is the diffusivity, positive on account of (\ref{4.9}).  The solution of (\ref{4.15}) may be written in the convolution form (\ref{3.8}).
In the case of small thermoelastic coupling we may regard $D_i$ as small so that the lower order wave is a travelling wave of permanent form $f(x-\hat{c}_it)$ modified by diffusion over a length scale $(D_it)^{1/2}$.  For small $\tau$, (\ref{4.11}) shows that the diffusivity (\ref{4.16}) becomes the same as the diffusivity (\ref{3.7}).  As we might expect, therefore, the lower order waves governed by (\ref{4.14}) become, for small $\tau$, governed instead by (\ref{3.6}), the equation for lower order waves in the classical case.  Equation (\ref{4.15}) has no counterpart in King {\em et al.} (1998) because the equation of telegraphy has no non-zero lower order wave speed.

To study the degenerate lower order wave operator $\partial_t$ in (\ref{4.10}) we replace it by zero everywhere else to obtain the diffusion equation 
\be\label{4.17}
\tau\overline{c}^2_4\cdot\frac{\overline{c}^2_1\overline{c}^2_2\overline{c}^2_3}
{\hat{c}^2_1\hat{c}^2_2\hat{c}^2_3}\,\partial^2_x\theta
 - \partial_t\theta = 0,
\ee
where an operator $\partial^6_x$ has been ignored.  For small $\tau$, use of (\ref{4.11}) shows that (\ref{4.17}) reduces to the diffusion equation (\ref{3.9}) of the zero-speed lower order wave of the classical case.

\subsection{The isotropic case}
For an isotropic material, in addition to (\ref{3.10}), we have 
\[\overline{c}^2_1=\overline{c}^2_2=\mu/\rho,\]
and the significant part of (\ref{4.9}) is
\[\hat{c}^2_2<\overline{c}^2_3<\tilde{c}^2_3<\hat{c}^2_3<\overline{c}^2_4.\]
Again, the transverse displacements $u_1$ and $u_2$  are  purely elastic in character and independent of temperature effects and satisfy the same isothermal wave equation as before.  Then the wave hierarchy form is obtained by replacing the lone operator $\partial_t$ commencing the second term of (\ref{3.11}) by $\Delta_t$ defined by (\ref{4.1}).  The effect of this is to show that the wave hierarchy form of the equations of isotropic generalized thermoelasticity may be obtained from (\ref{4.10}) by removing the common factors due to the two transverse isothermal waves to obtain
\be\label{4.18}
\tau\left(\partial^2_t-\overline{c}^2_3\partial^2_x\right)\left(\partial^2_t-\overline{c}^2_4\partial^2_x\right)\theta
+ \partial_t\left(\partial^2_t-\hat{c}^2_3\partial^2_x\right)\theta = 0.
\ee
After similarly removing the common factors from (\ref{4.8}), we find that
$\overline{c}^2_3$ and $\overline{c}^2_4$ may be obtained as the roots of the quadratic equation
\be\label{4.19}
v^2(v^2-\hat{c}^2_3) - (\rho c\tau)^{-1}k(v^2-\tilde{c}^2_3)=0.
\ee

By direct calculation, or by specializing the results of the previous subsection, the equations of disturbances propagating with the higher order wave speeds $\overline{c}_3$ and $\overline{c}_4$, with the lower order wave speed $\hat{c}_3$, and as the lower order degenerate diffusion operator are
\be\label{4.20}
\begin{array}{l}
\left(\partial_t + \overline{c}_3\partial_x\right)u_1 + \displaystyle\frac{\hat{c}^2_3 - \overline{c}^2_3}
{2\tau(\overline{c}^2_4-\overline{c}^2_3)}\, u_1=0, \\[4mm]
\left(\partial_t + \overline{c}_4\partial_x\right)u_1 + \displaystyle\frac{\overline{c}^2_3 - \hat{c}^2_3}
{2\tau(\overline{c}^2_4-\overline{c}^2_3)}\, u_1=0, \\[4mm]
\left(\partial_t + \hat{c}_3\partial_x\right)u_1 = 
\displaystyle\frac{(\hat{c}^2_3 - \overline{c}^2_3)(\overline{c}^2_4 - \hat{c}^2_3)}{2\hat{c}^2_3}
\partial^2_xu_1,\\[4mm]
\tau\overline{c}^2_4\cdot\displaystyle\frac{\overline{c}^2_3}
{\hat{c}^2_3}\,\partial^2_x\theta
 - \partial_t\theta = 0,
\end{array}
\ee
corresponding to (\ref{4.2}), $i=3, 4$, (\ref{4.15}) and (\ref{4.17}), respectively.  Longitudinal waves of sinusoidal form in generalized isotropic thermoelasticity were investigated by Leslie \& Scott (2000), who reached the same conclusions on stability.

\end{document}